\documentclass{aa}  
\usepackage{graphicx}
\usepackage{txfonts}
\newcommand{\etal}{et al.}
\newcommand{\ltsim}{\raisebox{-1mm}{$\stackrel{<}{\sim}$}}

\newcommand{\gtsim}{\raisebox{-1mm}{$\stackrel{>}{\sim}$}}
\def\arcm{\hbox{$^\prime$}}
\def\arcs{\arcm\hskip -0.1em\arcm}

\begin{document}

   \title{ XMM-Newton slew survey discovery of the nova XMMSL1~J070542.7-381442 (V598 Pup)}

   \author{A.\,M.~Read\inst{1}
           R.\,D.~Saxton\inst{2}
           M.\,A\,.P.~Torres\inst{3}
           P.~Esquej\inst{4}
           E.~Kuulkers\inst{2}
           P.\,G.~Jonker\inst{5,3}
           J.\,P.~Osborne\inst{1}
           M.\,J.~Freyberg\inst{4}
           P.~Challis\inst{3} 
           }


   \offprints{A.\,M.~Read}

   \institute{\inst{1}Dept.\ of Physics and Astronomy, Leicester University, Leicester LE1\,7RH, U.K.\\
              \email{amr30@star.le.ac.uk} \\
              \inst{2}ESA/ESAC, Apartado 78, 28691 Villanueva de la Ca\~{n}ada, Madrid, Spain\\
              \inst{3}Harvard--Smithsonian  Center for Astrophysics, Cambridge, MA~02138, U.S.A.\\
              \inst{4}Max-Planck-Institut f\"ur extraterrestrische Physik, 85748 Garching, Germany\\
              \inst{5}SRON, Netherlands Institute for Space Research, 3584~CA, Utrecht, The Netherlands \\
             }

   \titlerunning{XMM-Newton slew survey discovery of the nova XMMSL1~J070542.7-381442 (V598 Pup)}
   \authorrunning{A.\,M.~Read et al.}

   \date{Received September 15, 1996; accepted March 16, 1997}


  \abstract
{}
{In an attempt to catch new X-ray transients while they are still
  bright, the data taken by XMM-Newton as it slews between targets is
  being processed and cross-correlated with other X-ray observations
  as soon as the slew data appears in the XMM-Newton archive.}
{A bright source, XMMSL1~J070542.7-381442, was detected on 9 Oct
  2007 at a position where no previous X-ray source had been seen. The
  XMM slew data and optical data acquired with the Magellan Clay 6.5\,m telescope were
  used to classify the new object.}
{No XMM slew X-ray counts are detected above 1\,keV and the source is
  seen to be $\sim$750 times brighter than the ROSAT All-Sky Survey
  upper limit at that position. The normally m$_{V}$$\sim$16 star,
  USNO-A2.0 0450-03360039, which lies 3.5\arcs\ from the X-ray
  position, was seen in our Magellan data to be very much enhanced in
  brightness. Our optical spectrum showed emission lines which
  identified the source as a nova in the auroral phase.
  Hence this optical source is undoubtedly the progenitor of the
  X-ray source $-$ a new nova (now also known as V598 Pup).  The X-ray
  spectrum indicates that the nova was in a super-soft state (with
  $kT_{\rm eff}$$\approx$35\,eV). We estimate the distance to the nova to be $\sim$3\,kpc.  
  Analysis of archival robotic
  optical survey data shows a rapid decline light curve consistent with that 
  expected for a very fast nova.}
{The XMM-Newton slew data present a powerful opportunity to find new
  X-ray transient objects while they are still bright. Here we
  present the first such source discovered by the analysis of near
  real-time slew data.}
   
\keywords{Novae -- Stars: individual: V598 Pup -- Surveys -- X-rays: general}

   \maketitle
%

\section{Introduction}

To date, the publicly available XMM-Newton slew data covers over 25\%
of the sky, while the soft band (0.2$-$2 keV) slew sensitivity limit
(6$\times10^{-13}$\,ergs cm$^{-2}$ s$^{-1}$) is close to that of the
RASS (the hard-band [2$-$12 keV] limit is 4$\times10^{-12}$\,ergs
cm$^{-2}$ s$^{-1}$). For details of the slew data and catalogue, and
the first science results see Saxton \etal\ (2008) and Read \etal\
(2006). The near real-time comparison of XMM-Newton slew data with
ROSAT data is now giving, for the first time, the opportunity to find
all manner of high-variability X-ray objects, e.g.  tidal disruption
candidates (Esquej \etal\ 2007), AGN, blazars, and also Galactic
sources such as novae, flare stars, cataclysmic variables and
eclipsing X-ray binaries. It is only with such a large-area survey as
the XMM-Newton Slew Survey, that such rare events have a chance of
being caught $-$ within 
the first slew catalogue (XMMSL1, covering
14\% of the sky; Saxton \etal\ 2008), $\sim$40 individual slew sources are seen at fluxes $>$20
times greater than their corresponding ROSAT counterparts or 2$\sigma$
upper limits (assuming a 70\,eV black body model with Galactic $N_{\rm
  H}$). 55\% of these are believed to be new X-ray 
transients.

In an effort to find transient X-ray sources while they remain active,
we are now attempting to perform the slew data acquisition, analysis
and source-searching as quickly as possible. Catalogue
cross-correlations with RASS and ROSAT pointed data fluxes and upper
limits are swiftly preformed to identify highly variable X-ray
candidates. Slew datasets are made available in the XMM science
archive (XSA) typically $\sim$10 days after the slew has been
performed. This systematic processing of the most recent data has been
performed since October 2006. 

One such rare event,
XMMSL1~J070542.7-381442 was discovered in an XMM slew from 9 Oct 2007.
Here we describe the XMM-Newton slew observations, the identification
of the optical counterpart and a spectral confirmation of the source
as a nova in the auroral phase.  The optical lightcurve of the source
up to six months after outburst is also presented. Since the
discovery, follow-up observations have been made with XMM-Newton and
with Swift (Read \etal\ 2007b), and the X-ray flux is observed to
have declined continually. The discussion of these observations is
deferred to a later paper.


\section{XMM-Newton slew observations}

\begin{figure}
\centering
\includegraphics[bb=38 220 574 605,clip,width=8.0cm,angle=0]{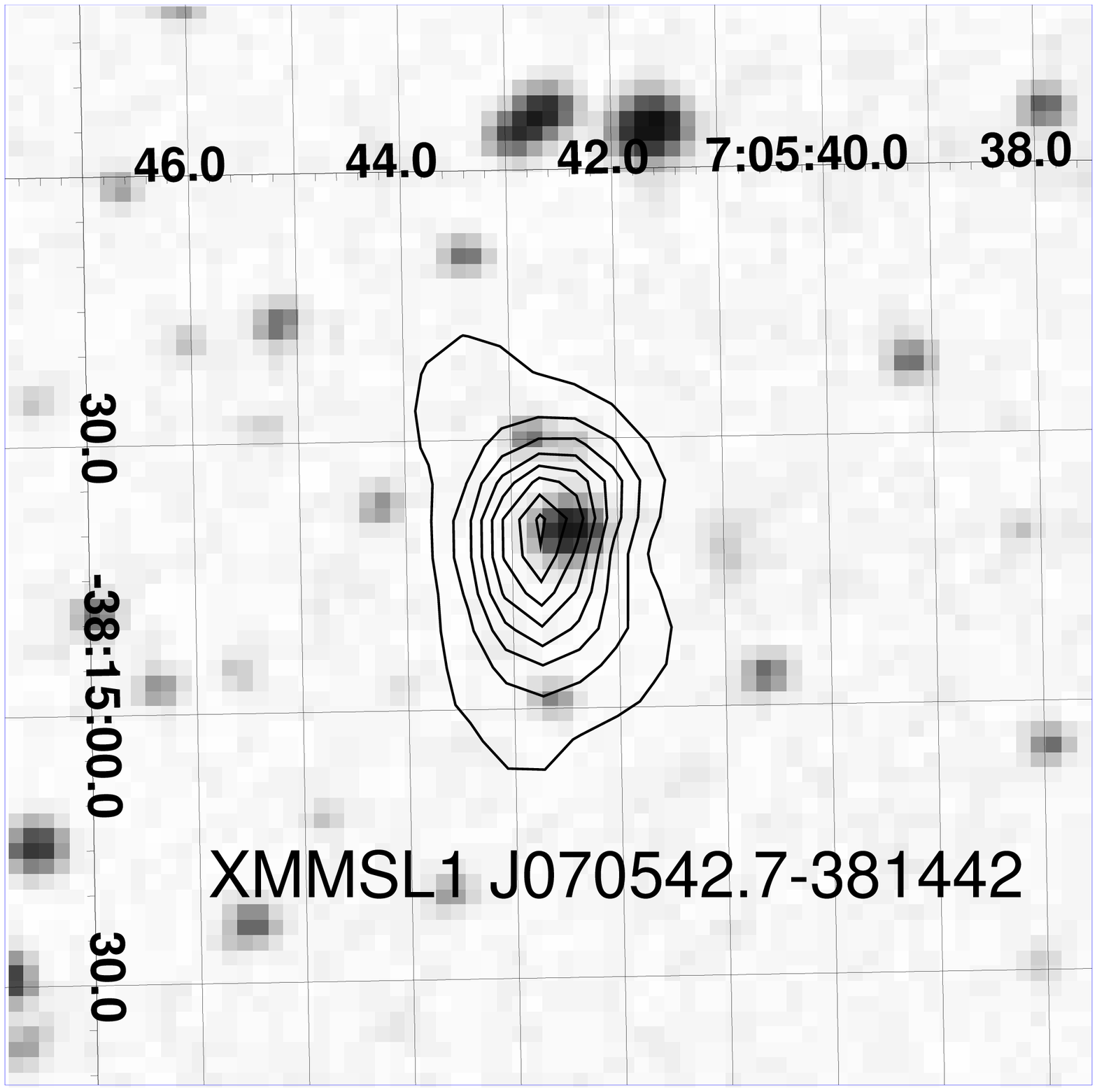}
\caption{Contours (16.8, 33.6, 50.4 etc. cts arcsec$^{-2}$) of
  (4\arcs\ FWHM) Gaussian-smoothed XMM-Newton slew data (EPIC-pn; 0.2-2\,keV) from
  XMMSL1~J070542.7-381442 superimposed on an optical DSS image. A
  slight elongation of the X-ray emission along the slew direction
  (due partly to the source slew drift during an individual EPIC-pn
  cycle time [73 ms] and also to the PSF shape at this off-axis
  position) is evident. The progenitor, the m$_{V}$$\sim$16 
  point-like source, USNO-A2.0 0450-03360039, is clearly visible.}
\label{slewim}
\end{figure}

XMMSL1~J070542.7-381442 was discovered at an X-ray position of
07:05:42.7 -38:14:42 (J2000; error radius: 8\arcs) in 
slew 9143400002 from XMM revolution 1434, made in the
EPIC-pn full-frame mode using the medium filter (Read \etal\
2007a). Contours of the XMM-Newton slew data are shown superimposed on
an optical DSS image in Fig.\ref{slewim}. This sky position has not
yet been observed during any other XMM-Newton slew.

The source passed through the centres of CCDs 3 \& 12 of the EPIC-pn
detector in 14\,s, at a large off-axis angle (minimum
$\approx$14\arcm), such that an effective vignetting-corrected soft
band (0.2$-$2\,keV) exposure time of 3.9\,s was achieved.  A total of
210 source counts lie within a radius of 20\arcs, yielding a (EPIC-pn:
0.2$-$2\,keV) count rate of 54.5\,ct s$^{-1}$, after correcting for
the encircled energy function.  The high count rate indicates that the
spectrum is affected by pile-up, though the effect here, far off-axis,
is less than on-axis (the on-axis limit is 6\,ct s$^{-1}$ for EPIC-pn
full-frame mode
\footnote{http://xmm.esac.esa.int/external/xmm\_user\_support/documentation
  /uhb\_2.5/index.html}). X-ray loading (Smith 2004), where events
below the cut-off threshold sum together to produce artificial
accepted events, is also present. These effects work to distort the
spectrum and make quantitative spectral analysis difficult. One can
minimize these effects by ignoring the central part of the Point
Spread Function, and we therefore extracted a spectrum of the source
from within an annulus of 5\arcs$-$60\arcs\ radius, centred on the
source position.  Just single events were selected, and these were
spectrally grouped to give a minimum of 20\,counts per bin. Slew data
also has as yet unresolved problems associated with the motion of the
source across the detector: approximations currently have to be made
when calculating the associated effective area and detector response
matrix files. To perform a qualitative spectral analysis, an effective
area file, accounting for the removal of the piled-up core, was
generated by averaging the individual core-removed effective area
files at 9 different positions along the detector track made by the
source. This takes into account to a good approximation the variations
in the vignetting and the PSF. Individual BACKSCAL values have been
set by hand, as have the EXPOSURE values, estimated by calculating the
distance travelled by the source in detector coordinates and finding
the time taken to do this given a 90\,deg\,hr$^{-1}$ slew speed, then
subtracting the appropriate fractions for chip gaps and bad pixels
(calculating the exposure time from the source lightcurve, gives the
same value to within a few tenths of a second). For the response
matrix, we used the equivalent canned detector response matrix for the
vignetting-weighted average source position, for single events and for
Full Frame mode: epn\_ff20\_sY6\_v6.9.rmf. A background spectrum was
extracted from a much larger region close to the source and at a
similar off-axis angle.

\begin{figure}
\centering
\includegraphics[bb=110 20 580 750,clip,width=6.0cm,angle=270]{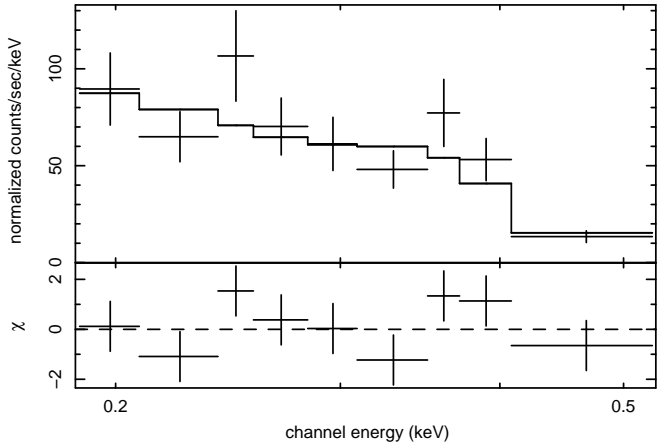}
\caption{The XMM-Newton slew spectrum of XMMSL1~J070542.7-381442, fit
  with a white dwarf atmosphere model of effective temperature 35\,eV.
  The data and model are shown in the upper panel and the deviations
  of the data from the model are shown in the lower panel.}
\label{slewspec}
\end{figure}

Simple power-law, blackbody, thermal Bremmstrahlung and optically thin
hot plasma models are unable to fit the spectrum adequately (all have
a $\chi^{2}$$\sim$12 for 6 degrees of freedom). Given that the source
is identified as a nova (Section~3), a more physically realistic white
dwarf atmosphere model, of the type used to model the nova V1974 Cyg
(Balman \etal\ 1998), was used, yielding an acceptable fit (reduced
$\chi^{2}$$\approx$1.4 for 6 degrees of freedom), an effective
temperature of 35$^{+2}_{-1}$\,eV and an $N_{\rm H}$ of
4.8$^{+5.7}_{-2.7}$$\times$$10^{20}$\,cm$^{-2}$ (errors 90\% for one
interesting parameter). The spectrum and model fit are shown in
Fig.\ref{slewspec}.  The model normalization, corrected for the
removal of the saturated PSF core, can be used to derive a distance to
the source (assuming the emission to be sub-Eddington, and a typical
emitting region of spherical radius 10$^{9}$\,cm) of
$\approx$4.3$\pm2.3$\,kpc. A
PIMMS\footnote{http://heasarc.nasa.gov/Tools/w3pimms.html} v3.9d
conversion of the RASS 0.0076\,ct s$^{-1}$ 2$\sigma$ upper limit
corresponds to an EPIC-pn limit $\sim$750 times less than observed of
0.073\,ct s$^{-1}$, assuming the same spectral model.

\section{Optical observations}

The field containing XMMSL1~J070542.7-381442 was observed on 16 Nov
2007 with the Magellan Clay 6.5m telescope at Las Campanas
Observatory. A bright, saturated optical counterpart was found with a
position consistent with the m$_{V}$$\sim$16 point-like source
USNO-A2.0~0450-03360039 (07:05:42.5 
\linebreak
-38:14:39; J2000), 3.5\arcs\ from
the slew position. On the basis of the positional coincidences, the
USNO object was proposed as a possible progenitor to the nova (Read
\etal\ 2007a, Torres \etal\ 2007). This claim was confirmed with the
detection of the radio counterpart to the slew source, which has a
high-precision position consistent with the USNO object (Rupen \etal\
2007).

Spectroscopic observations of the optical counterpart were obtained
with the Low Dispersion Survey Spectrograph (LDSS-3), equipped with
the 400 line~mm$^{-1}$ VPH ALL grism and a mask with a 1\arcs\ width
long-slit cut near the centre of the field of view. The detector was
the STA0500A 4k$\times$4k unbinned CCD. This setup allowed us to cover
the spectral range 3510--10620~\AA\ with a dispersion of
2.0~\AA~pix$^{-1}$ and a resolution of $\approx$8~\AA~FWHM.  Several
exposures (2$\times$1\,s, 1$ \times$2\,s, 1$\times$5\,s and
1$\times$60\,s) were obtained.  These spectra were affected by
2nd-order light contamination beyond $\sim$7000~\AA. In order to
obtain useful coverage in the red, we also acquired 1\,s, 5\,s and
60\,s spectra with the OG590 order-blocking filter, giving useful
5800--10620~\AA\ wavelength coverage.  The frames were reduced using
standard routines in IRAF. The spectra were then extracted and
wavelength-calibrated with the help of HeNeAr calibration lamp
spectra.  The instrumental response was corrected using spectroscopic
standard stars observed with and without the blocking filter. The
shape of the continuum is reliable except at the blue and red ends.

\begin{figure}
\centering
\includegraphics[bb=18 178 566 692,clip,width=8.0cm,angle=0]{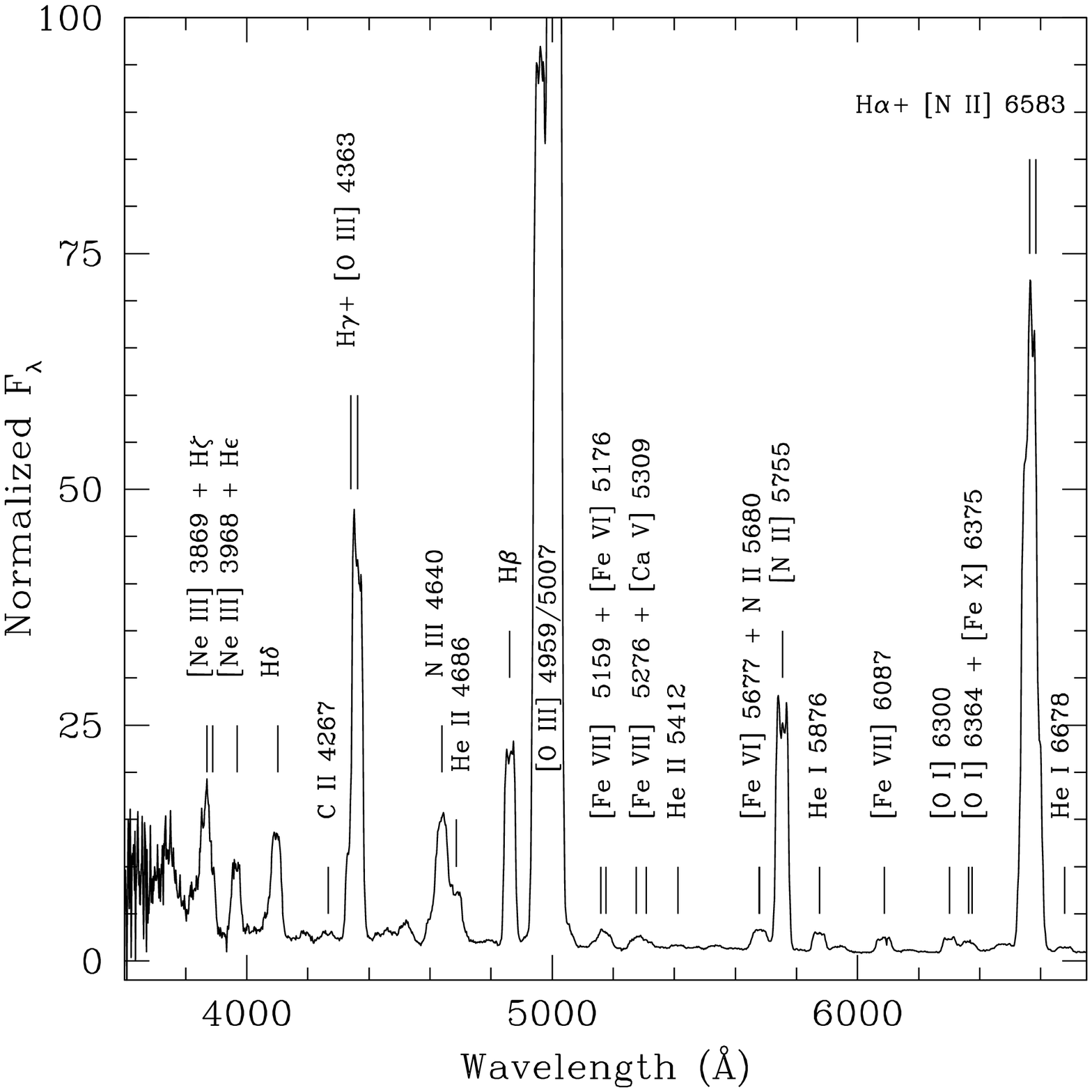}
\includegraphics[bb=33 153 566 687,clip,width=8.0cm,angle=0]{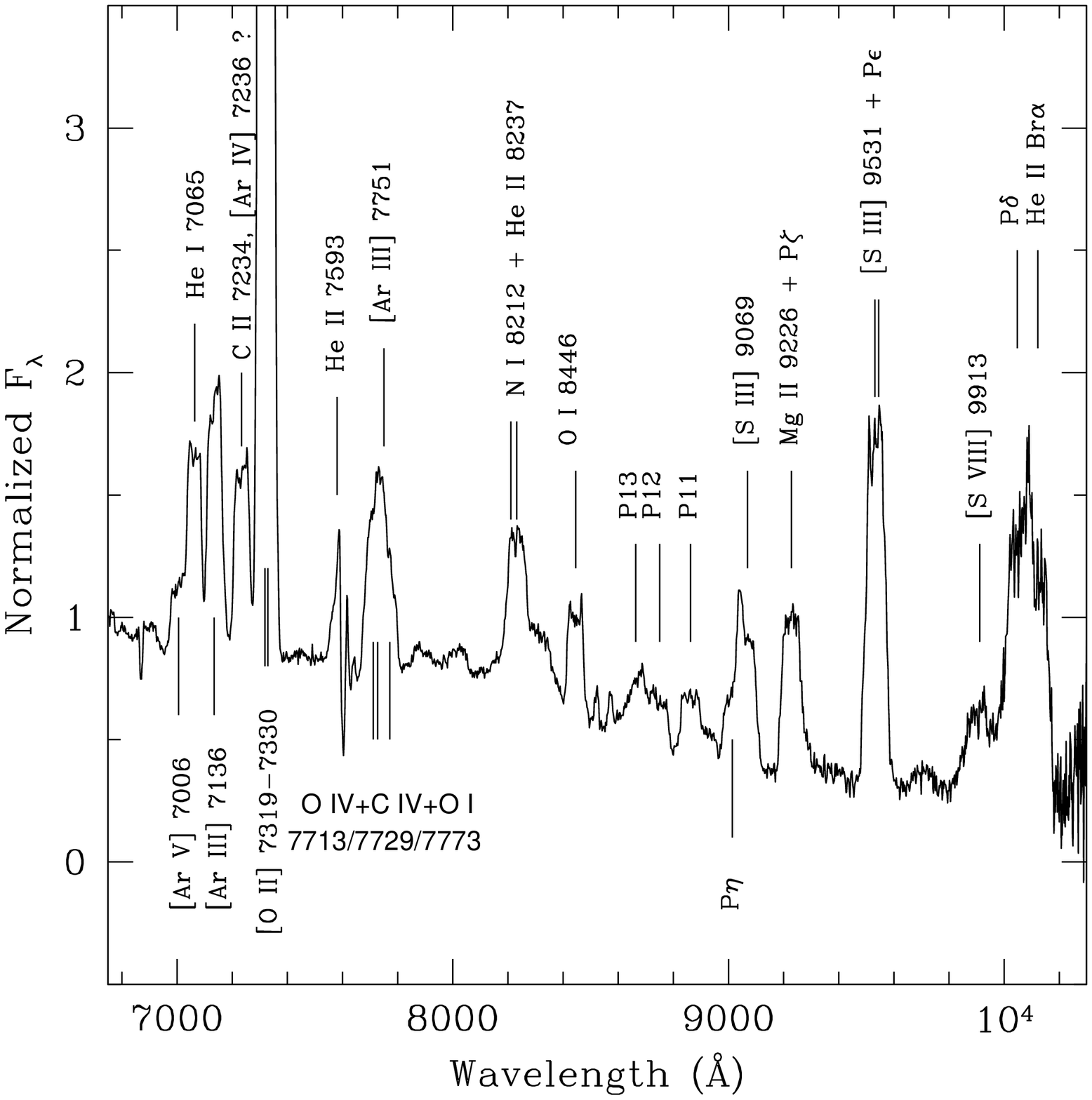}
\caption{Blue (top) and red (bottom) spectra of the optical counterpart to
XMMSL1~J070542.7-381442 acquired on 16/11/07. The spectra have been
normalized to unity at $\lambda$6750. Stronger emission lines are identified. The
absorption feature at He {\sc ii} $\lambda$7593 is telluric in origin.}
\label{magspec}
\end{figure}

A blue and a red spectrum are shown in Fig.\ref{magspec}. The spectra
were obtained 163 days after the maximum optical brightness reported
by Pojmanski \etal\ (2007) (see Section~4).  The data show emission
lines characteristic of a late post-outburst nova spectrum. The
strongest lines in the blue spectrum are [0 {\sc iii}] $\lambda
5007,4959$ followed by H$\alpha$ blended with [N {\sc ii}] $\lambda
6583$, [O {\sc iii}] $\lambda 4363$ (blended with H$\gamma$), [N {\sc
  ii}] $\lambda 5755$ and H$\beta$.  The strongest emission feature in
the red spectrum is due to [O {\sc ii}] $\lambda\lambda 7319$--$7330$
auroral transtions.  The emission lines have an average FWHM of $2070
\pm 50$ km~s$^{-1}$, as derived from isolated line profiles and
correcting for the instrumental broadening. H$\beta$ is stronger than
He {\sc ii} $\lambda 4686$. This characteristic together with the
presence of strong forbidden [O {\sc iii}], [N {\sc ii}] and [O {\sc
  ii}] auroral lines, and weak Ne lines suggest that this source was a
Fe {\sc ii} nova (see e.g.  Williams 1992). However, the data were
acquired too long after maximum light to make an accurate
classification possible.  The nova was observed in the A$_{0}$ auroral
phase according to the Tololo classification (Williams et al. 1991;
Williams \etal\ 1994) $-$ any forbidden auroral transition at
wavelengths $\lambda\lambda 3600$--$7600$ has a flux larger than that
of the strongest non-Balmer permitted lines, and [O {\sc iii}]
$\lambda 4363$ is the strongest auroral transition. The coronal [Fe
{\sc x}] $\lambda 6375$ line (if present) is weaker than [Fe {\sc
  vii}] $\lambda 6375$, excluding the possibility of a coronal stage.

\section{Optical light curve}

\begin{figure}
\centering
\includegraphics[bb=55 78 453 549,clip,width=7.0cm,angle=270]{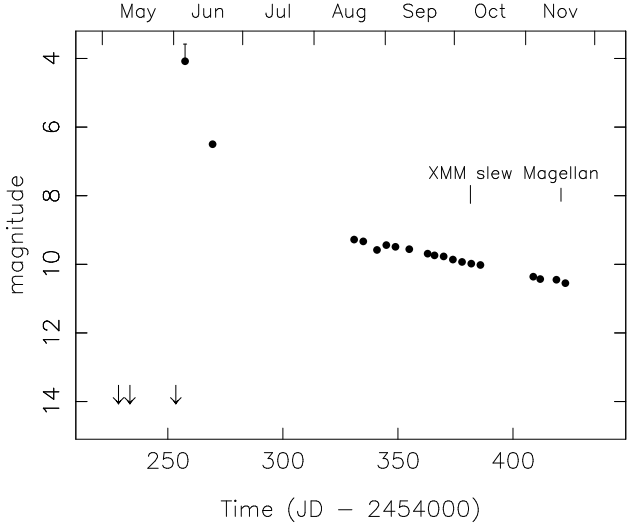}
\caption{V-band magnitudes from Pojmanski \etal\ (2007) of the optical
  counterpart to XMMSL1~J070542.7-381442. The 2007 dates of the
  XMM-Newton slew and Magellan observations are marked.}
\label{lc}
\end{figure}

Analysis of archival robotic optical survey data from 3-minute CCD
exposures (pixel size 14\arcs.8), obtained with a 70\,mm (200\,mm
focal length) f/2.8 telephoto lens in the course of the All Sky
Automated Survey (Pojmanski 2002) show that the visual magnitude of
this source rose from m$_{V}\gtsim$14 to m$_{V}$$\approx$4.1 between 2
Jun (23:27\,UT) and 5 Jun (23:13\,UT) 2007, and has declined since
(see Fig.\ref{lc}). The source was seen to be saturated in the June
5th observation, and it is thought that the source may have been
0.1$-$0.5 magnitudes brighter.  The decline from outburst ($\gtsim$2.4
magnitudes in 12 days, then a further 2.8 magnitudes in 62 days)
indicates that this is a nova of speed class 'very fast' (Warner
1995). We estimate $t_{2}$, the time that the light curve takes to
decline 2 magnitudes below maximum brightness as 9$\pm$1\,days. The
later decay rate (mid-August onwards) is 1.36$\pm$0.04\,mag per 100
days (including errors to obtain a reduced $\chi^{2}$ of $\approx$1
for the fit).

\section{Discussion}

XMMSL1~J070542.7-381442, as a confirmed classical nova, has been given
the name V598 Puppis (Samus 2007). 
With a peak $m_{V}$$\ltsim4$, it is one of the brightest
optical novae seen for many years. A 
$m_{V}$$\ltsim4$ nova has been discovered in eruption every $\approx$8 years or so (Warner 1995), 
and the only novae discovered in recent decades of comparable peak optical magnitude 
are V382 Vel ($m\sim3$) and V1494 Aql
($m\sim4$), both discovered in 1999, and V1280 Sco ($m\sim4$),
discovered in Feb 2007.

Classical novae
are usually discovered optically in the early phases
of the outburst, as they are intrinsically optically bright and easily
found in inexpensive wide-area shallow surveys. Unusually in this
case, the optical outburst went unnoticed, and V598~Pup was only
discovered in X-rays during the later ($\sim$100 days after outburst),
optically thin nebular phase, when classical novae are typically
observed as soft X-ray sources.

On the basis of the optical spectrum, this nova was observed in the
A$_{0}$ auroral phase, and was likely a very fast, Fe~{\sc ii} nova
(Section~3 and Williams et al. 1991; Williams \etal\ 1994), though an
accurate classification is now, late after maximum brightness, not
possible. The soft ($kT_{\rm eff}$=35\,eV) X-ray spectrum
indicates that the nova was in a super-soft state (Pietsch \etal\
2007). Such a state is due to nuclear burning on the white dwarf (Ness
\etal\ 2007). Measurement of its intensity, duration and temperature
can constrain the distance to the nova and the mass of the white dwarf
(e.g. Balman \etal\ 1998; Lanz \etal\ 2005). From the slew measurement
of V598 Pup, we see that the delay from the outburst (2$-$5 Jun
2007) to the onset of the X-ray super-soft state is
$\ltsim$127 days.  This is short when compared with the $\sim$200 days
seen in V1974 Cyg (Krautter \etal\ 1996), $\sim$6 months of V382 Vel
(Orio \etal\ 2002) and 6$-$8 months of V1494 Aql (Drake \etal\ 2003).

From the Galactic latitude and the fact that the Galactic scale height
of white dwarfs is $<$500\,pc (conservatively, Nelson \etal\ 2002), an
upper limit to the distance of V598 Pup of $\approx$2.1\,kpc can be
derived, consistent with the 4.3$\pm2.3$\,kpc, estimated from the
X-ray spectral fitting. Another way to estimate the distance is to use
the relation between absolute magnitude at maximum brightness and $t_{2}$
(see e.g.\,equation~5.2 in Warner 1995).
Using $t_{2}$=9$\pm$1\,days, we estimate $M_{V}$=-8.4$\pm$0.4. With
$A_{V}$=0.27$^{+0.31}_{-0.15}$ (90\% error), as derived (Predehl \& Schmitt 1995) 
from $N_{\rm H}$=4.8$^{+5.7}_{-2.7}$$\times$$10^{20}$\,cm$^{-2}$, 
$M_{V}$=-8.4$\pm$0.4 and peak $m_{V}$=4.1, we derive a distance of
2.8$^{+0.8}_{-0.5}$\,kpc. 
An absolute magnitude of $M_{V}$=-8.4 would imply a
peak luminosity of order $\sim$7 times the Eddington luminosity for a
1\,$M_{\odot}$ white dwarf. This is quite typical for novae.
The source had, at the time of the slew detection, an absorbed
(0.2$-$2\,keV) X-ray flux of 1.54$^{+0.08}_{-0.23}\times10^{-10}$\,erg
cm$^{-2}$ s$^{-1}$, corresponding to, assuming a distance of 3\,kpc, a
(0.2$-$2\,keV) X-ray luminosity of
1.6$^{+0.1}_{-0.2}$$\times$$10^{35}$\,erg s$^{-1}$, and a bolometric
luminosity of 2.4$^{+6.4}_{-1.7}$$\times$$10^{36}$\,erg s$^{-1}$ (errors
calculated at the boundary of the 90\% region for two interesting
parameters; $N_{\rm H}$ \& $kT_{\rm eff}$). This is at the lower end of 
the luminosities 
discussed e.g. in Orio \etal\ (2002) and Ness \etal\ (2007). 


\begin{acknowledgements}

  This research made use of the VIZIER database, operated at CDS,
  Strasbourg, France. The XMM-Newton project is an ESA Science Mission
  with instruments and contributions directly funded by ESA Member
  States and the USA (NASA). 
  We thank K.\,L.\,Page \& M.\,Modjaz for useful discussions, and the
  referee for useful comments which have improved the paper. AMR \&
  JPO acknowledge the funding support of PPARC/STFC, PE of MPE, and
  PGJ of the Netherlands Organisation for Scientific Research.

\end{acknowledgements}

\end{document}